\documentclass[aps,twocolumn,prd,nofootinbib,preprintnumbers,amsmath,amssymb,superscriptaddress]{revtex4}
\makeatletter
\def\section{%
  \@startsection
    {section}%
    {1}%
    {\z@}%
    {0.8cm \@plus1ex \@minus .2ex}%
    {.2cm} 
    {\normalfont\small\bfseries}%
}%
\def\subsection{%
  \@startsection
    {subsection}%
    {2}%
    {\z@}%
    {.8cm \@plus1ex \@minus .2ex}%
    {.2cm}%
    {\centering\normalfont\small\bfseries}%
}%
\makeatother

\usepackage{graphicx}
\pdfoutput=1

\usepackage{hyperref}
\usepackage{mathtools}
\usepackage{amssymb, amsmath, amsopn, amsthm}
\usepackage{epsfig,amsfonts,bm}
\usepackage{graphics}
\usepackage{comment}

\allowdisplaybreaks[1]

\newcommand{\tbref}[1]{{\tablename\;\,\ref{#1}}}
\newcommand{\fgref}[1]{{\figurename\,\ref{#1}}}

\def\l{\left}
\def\r{\right}
\newcommand{\nl}{\nonumber\\*}

\newcommand{\abs}[1]{{\left|#1\right|}}

\newcommand{\ti}[1]{\tilde{#1}}
\newcommand{\f}[2]{{\frac{#1}{#2}}}
\newcommand{\s}[1]{\sqrt{#1}}

\newcommand{\bR}{\mathbb{R}}

\newcommand{\bZ}{\mathbb{Z}}

\newcommand{\SO}{\mathrm{SO}}

\newcommand{\SL}{\mathrm{SL}}

\begin{document}

\title{On the Entanglement of Multiple CFTs \\ via Rotating Black Hole Interior}
\author{Norihiro Iizuka}
\email{norihiro.iizuka@riken.jp}
\thanks{\\Address after April 1, 2014: {\itshape\footnotesize{Department of Physics, \\Osaka University, Toyonaka, Osaka 560-0043, JAPAN}}}
\affiliation{Interdisciplinary Fundamental Physics Team, 
Interdisciplinary Theoretical Science Research Group,  RIKEN, Wako 351-0198, JAPAN} 
\author{Noriaki Ogawa}
\email{noriaki@kias.re.kr}
\affiliation{School of Physics, Korea Institute for Advanced Study, 
Seoul 130-722, Korea}


\begin{abstract}
We study the minimal surfaces between two of the multiple boundaries of 3d maximally extended rotating eternal black hole. Via AdS/CFT, this corresponds to investigating the behavior of entanglements of the boundary CFT with multiple sectors. Non-trivial time evolutions of such entanglements detect the geometry inside the horizon, and behave differently depending on the choice of the two boundaries.
\end{abstract}

\preprint{KIAS-P14006}


\maketitle



\section{Introduction}\label{sec:intro}

The gauge/gravity correspondence is a fascinating correspondence. 
In short, it gives the non-perturbative definition of quantum gravity in terms of the corresponding gauge theories.  
However, how and why the bulk spacetime and gravity appears out of the boundary theory is still in obscurity, 
and indeed it is one of the most fundamental questions in this correspondence.

Van Raamsdonk \cite{VanRaamsdonk:2009ar,VanRaamsdonk:2010pw}
pointed out that quantum entanglement between separated regions of the boundary theory 
is a key to the emergence of smoothly connectedness of the spacetime in the bulk. 
This idea has been extended and materialized in subsequent works,
for example
\cite{Hartman:2013qma,Maldacena:2013xja}.
Especially in Hartman-Maldacena \cite{Hartman:2013qma}, they investigated the time evolution of entanglement between the two copies of boundary CFT, in eternal AdS black hole
without angular momentum.
In this letter,
we extend the analysis of \cite{Hartman:2013qma} 
to the case of rotating BTZ black hole.
Unlike the non-rotating case, the spacetime boundary has $8$ disconnected regions. Naively they seem to correspond to $8$ decoupled sectors of the boundary CFT, which are (maximally) entangled to one another.
However, as we will see, actually the story is not so simple. 

This short letter is organized as follows. 
We first review the known global structure of the rotating BTZ black hole
very briefly in section \ref{sec:BTZ} (a bit more details are given in Appendices \ref{app:BTZ},\ref{app:analytic}).
Then, we derive the lengths of the geodesics connecting the different boundaries
in section \ref{sec:geodesics},
and by using it, calculate the entanglement between different boundaries in section \ref{sec:EE} using the 
holographic entanglement formula \cite{Ryu:2006bv,Ryu:2006ef}. 
We will discuss and interpret the results in section \ref{sec:discussion}.

\section{Rotating BTZ and Analytic Continuation}
\label{sec:BTZ}
The rotating BTZ black hole geometry is expressed as
\begin{align}
\label{eq:BTZmetric}
  ds^2 &= -f(r)dt^2 + \frac{dr^2}{f(r)} + r^2\l(d\phi  - N(r)dt\r)^2\,,\\
  f(r) &= \f{(r^2-r_+^2)(r^2-r_-^2)}{r^2}\,,
\quad
  N(r) = \f{r_+r_-}{r^2}\,, 
  \label{eq:phiperiod}
\end{align}  
where  $ \phi \simeq \phi+2\pi$. We set AdS scale to be unit in this letter. 
The outer/inner horizon radii $r_+$ and $r_-$ are related to 
the 
``chiral temperatures'' $T_+$ and $T_-$ ($T_+ \le T_-$)
as $r_{\pm} = \pi  (T_-\pm T_+)$.

This geometry is obtained by an orbifold on the global AdS$_3$,
and the outer region of the horizon ($r>r_+$)
can be embedded in $\bR^{2,2}$,
where $ds^2=-dx_0^2-dx_1^2+dx_2^2+dx_3^2$,
as
\begin{subequations}
\begin{align}
  x_1 &= \eta_1\l(\f{r^2-r_-^2}{r_+^2-r_-^2}\r)^{1/2}\cosh(\pi\l(T_+u^++T_-u^-\r))\,,\\
  x_2 &= \eta_1\l(\f{r^2-r_-^2}{r_+^2-r_-^2}\r)^{1/2}\sinh(\pi\l(T_+u^++T_-u^-\r))\,,\\
  x_3 &= \eta_2\l(\f{r^2-r_+^2}{r_+^2-r_-^2}\r)^{1/2}\cosh(\pi\l(T_+u^+-T_-u^-\r))\,,\\
  x_0 &= \eta_2\l(\f{r^2-r_+^2}{r_+^2-r_-^2}\r)^{1/2}\sinh(\pi\l(T_+u^+-T_-u^-\r))\,,
\end{align}
\end{subequations}
where $u^{\pm}=\phi\pm t$ and $\eta_i=\pm 1$.
The four combinations of the $(\eta_1,\eta_2)$
represents distinct regions outside the black hole, 
which we call $1_{++}$, $1_{+-}$, $1_{-+}$ and $1_{--}$.\
One can go from one to another only through the interior region.
We explain the spacetime structure of this geometry slightly more in Appendix \ref{app:BTZ}. 
The orbifold to produce the periodicity for $\phi$ is given in \eqref{eq:orbifold}. 
For more details, see \cite{Hemming:2002kd}.

Furthermore, these different regions 
can be connected to one another by analytic continuations of $(t,\phi)$ or $u^{\pm}$ coordinates to complex-valued regions, as \tbref{tab:analytic1}.

\begin{table}
  \centering
  \begin{tabular}{c|c|c|c}
     & $u^+$ & $u^-$ & \;$r$\;\\
\hline
    $1_{++}$ & $u^+$ & $u^-$ & $r$\\
    $1_{--}$ & $u^+-\f{i}{T_+}$ & $u^-$ & $r$\\
    $1_{+-}$ & $u^+-\f{i}{2T_+}$ & $u^-+\f{i}{2T_-}$ & $r$\\
    $1_{-+}$ & $u^+-\f{i}{2T_+}$ & $u^--\f{i}{2T_-}$ & $r$
  \end{tabular}
  \caption{Analytic continuations from $1_{++}$ to $1_{\eta_1\eta_2}$,
up to the periodicity $(u^+,u^-)\simeq (u^++ i/T_+, u^-\pm i/T_-)$. 
Their complex conjugates also work. 
}
  \label{tab:analytic1}
\end{table}

\section{Geodesics between Boundaries}
\label{sec:geodesics}
Our purpose in this paper is to investigate the way how the degrees of freedom on
different boundaries are entangled. 
In 2d CFT, 
various entanglement entropies
are expressed as combinations of the lengths of geodesics connecting points on the boundaries of the 3d spacetime, according to 
the Ryu-Takayanagi holographic entanglement entropy formula \cite{Ryu:2006bv,Ryu:2006ef}.

First, we consider two points $P_1=(t_1,\phi_1,r_{\infty})$
and $P_2=(t_2,\phi_2,r_{\infty})$ on the boundary of the same region,
say $1_{++}$.
The geodesic length connecting $P_1$ and $P_2$ is easily computed
by the coordinate mapping to Poincare AdS$_3$ \cite{Hubeny:2007xt}, giving
\begin{align}
  \label{eq:geodesic1++1++}
  &L_{1_{++}}^{(n)}(P_1,P_2)= 
\log{X_n}
-\log{\l(\pi^2T_+T_-\r)} + \log{r_{\infty}^2}\,,\nonumber\\
  &X_n = \sinh(\pi T_-(\delta u^-+2n\pi))
\sinh(\pi T_+(\delta u^++2n\pi))\,,\nonumber\\
  &\delta u^{\pm} = u_2^{\pm} - u_1^{\pm}\,,
\end{align}
where $n\in\bZ$ is the ``winding number'' around the $\phi$-circle \eqref{eq:phiperiod}.
The minimum $X=\min_{n\in\bZ}\{X_n\}$ is positive, 
if and only if $P_1$ and $P_2$ are spacelikely separated on the cylindrical boundary of $1_{++}$.

By applying the analytic continuation \tbref{tab:analytic1}
for the point $P_2$ in \eqref{eq:geodesic1++1++},
we obtain the geodesic length between $1_{++}$ and another boundary,
as \tbref{tab:geodesics}.

\begin{table}
\center
\begin{tabular}{l|r}
 & \multicolumn{1}{c}{$X_n$} \\
\hline
$1_{++}$ & $\sinh(\pi T_-(\delta u^-+2\pi n))\sinh(\pi T_+(\delta u^++2\pi n))$ \\
$1_{--}$ & $-\sinh(\pi T_-(\delta u^-+2\pi n))\sinh(\pi T_+(\delta u^++2\pi n))$ \\
$1_{+-}$ & $\cosh(\pi T_-(\delta u^-+2\pi n))\cosh(\pi T_+(\delta u^++2\pi n))$ \\
$1_{-+}$ & $-\cosh(\pi T_-(\delta u^-+2\pi n))\cosh(\pi T_+(\delta u^++2\pi n))$
\end{tabular}

\caption{Lengths $L^{(n)}(P_1,P_2)$ of geodesics connecting $P_1$ on $1_{++}$ boundary and $P_2$ on each boundary, 
in terms of $X_n$ where 
$L^{(n)}(P_1,P_2) = \log{X_n} -\log{\l(\pi^2T_+T_-\r)} + \log{r_{\infty}^2}$
and $\delta u^{\pm} = u_2^{\pm} - u_1^{\pm}$.
}
\label{tab:geodesics}
\end{table}

From this \tbref{tab:geodesics},
we notice that $X_n$ are always positive for $1_{+-}$
and negative for $1_{-+}$.
It implies that whole of the $1_{+-}$ boundary is spacelikely separated
to $1_{++}$ boundary, while the $1_{-+}$ boundary is timelikely.
The most complicated is the case of $1_{--}$.
By taking very large winding number $n$, we can make $X_n$ arbitrarily negative, that is, make the geodesic more timelike.

\section{Entanglement between different boundaries}
\label{sec:EE}
When we take a proper time-slice in this spacetime which connects the boundaries of $1_{++}$ and $1_{+-}$, the boundary dual is discussed in \cite{Maldacena:2001kr} ---
it is a maximally entangled pair of two CFT sectors with chiral temperatures $T_+\ne T_-$ (in other words, with chemical potential for momentum).
We expect that it would also be true when we take different time-slices connecting other boundary pairs.

\begin{figure}
  \centering
  \includegraphics[width=0.4\textwidth]{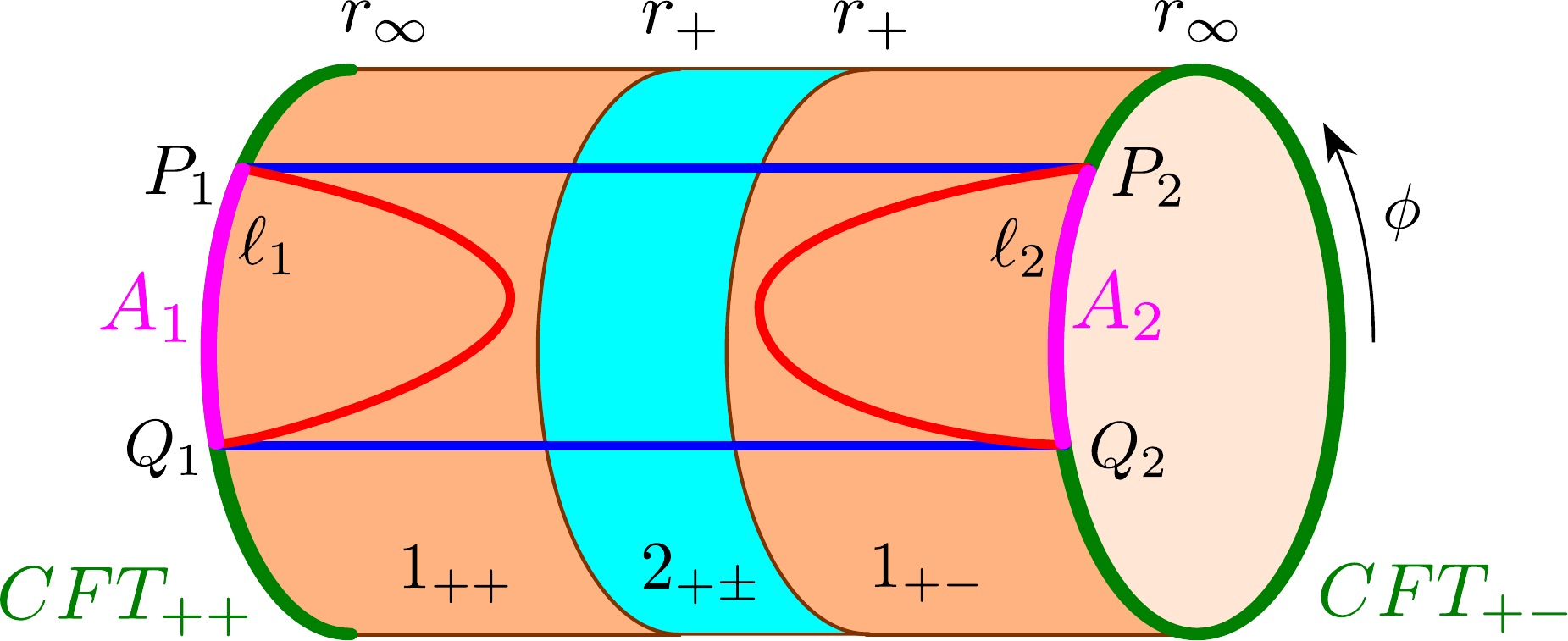}
  \caption{The subsystem $A=A_1\cup A_2$ and the corresponding 
candidates of the minimal surface, when $A_2$ is on the boundary of $1_{+-}$.
Red lines: the ``disconnected'' surface (eq.\eqref{eq:SAd}).
Blue lines: one of the ``connected'' surfaces across the horizon
(eq.\eqref{eq:SAc}, $n=0$).
Although the time direction is not drawn here, 
those lines do not live on the same time slice in general.
}
  \label{fig:EEsetup}
\end{figure}

In order to investigate the structures of such inter-boundary entanglement,
we consider the entanglement entropy for 
a subsystem $A$ which is the union of two intervals $A_1$ and $A_2$
on different boundaries (\fgref{fig:EEsetup}).
We fix $A_1$ on the boundary of $1_{++}$,
and $A_2$ is on another one
(in \fgref{fig:EEsetup}, $1_{+-}$).

We set the endpoints of $A_1$ and $A_2$ as
 $P_1=(t_1,\phi_1,r_\infty)$, $Q_1=(t_1,\phi_1+\ell_1,r_\infty)$
and $P_2=(t_2,\phi_2,r_\infty)$, $Q_2=(t_2,\phi_2+\ell_2,r_\infty)$,
respectively,
where $0<\ell_1, \ell_2 < 2\pi$.

According to the minimal area prescription \cite{Ryu:2006bv,Ryu:2006ef},
the corresponding entanglement entropy is given by
\begin{align}
  \label{eq:SA}
  S_A&= \min\l\{S_A^{(c)}\,, S_A^{(d)}\r\}\,,\\
  \label{eq:SAd}
  S_A^{(d)} &= L^{(0)}(P_1,Q_1) + L^{(0)}(P_2,Q_2)\,,\\
  \label{eq:SAc}
  S_A^{(c)} &= \min\l\{\l.L^{(n)}(P_1,P_2) + L^{(n)}(Q_1,Q_2)\r|n\in\bZ\r\}\,,
\end{align}
\footnote{Actually $S_A^{(d)}$ has another candidate which corresponds to the surface going around the other side of the $\phi$-circle (with winding number $n=-1$).
Hereafter we assume that \eqref{eq:SAd} (red line in \fgref{fig:EEsetup}) is always smaller than it. 
This is possible without loss of generality 
because we can redefine $A_i\to A_i^c$ and $\ell_i\to 2\pi -\ell_i$  ($i=1,2$ at the same time) without changing $S_A^{(c)}$ \eqref{eq:SAc}.
}
where we set $4G_N=1$ for simplicity.
These $S_A^{(d)}$ and $S_A^{(c)}$ correspond to different topologies
of the minimal surface drawn in \fgref{fig:EEsetup}
by red (``disconnected'') and blue (``connected'') lines%
\footnote{In the connected phase,
the two geodesics in \eqref{eq:SAc} must have same winding numbers, 
in order that the union of the two geodesics should 
be homotopic to $A$.
}. 
Physical quantity 
\begin{align}
  I(A_1,A_2) = S_{A_1} + S_{A_2} - S_A\,,
\end{align}
plays the role of the order parameter which distinguishes these two phases. 
That is, the red disconnected surface corresponds to $I(A_1,A_2)=0$ 
phase while the blue connected one is $I(A_1,A_2)>0$ phase, and 
it is a sharp phase transition only in the classical approximation, i.e., large $N$
on the CFT side \cite{Headrick:2010zt,Hartman:2013qma}\footnote{The authors thank J. Maldacena for explaining this point.}.

The entanglement entropy of the disconnected phase
$S_A^{(d)}$ \eqref{eq:SAd}
can be written as
\begin{align}
\label{eq:SAd_result}
  S_A^{(d)} =& 
\log[\sinh(\pi T_-\ell_1)\sinh(\pi T_+\ell_1)
\sinh(\pi T_-\ell_2)\sinh(\pi T_+\ell_2)] 
\nl
&- 2\log(\pi^2T_+T_-) + 2\log{r_{\infty}^2}\,,
\end{align}
regardless of which boundary $A_2$ lives on.
In particular, when the black hole is nearly extremal,
we have $T_+^{-1}\gg \ell_i$ and then
\begin{align}
  S_A^{(d)}
\simeq\,&
  \log[\sinh(\pi T_-\ell_1)\sinh(\pi T_-\ell_2)] + \log (\ell_1 \ell_2)
\nl
  &-2\log{(\pi T_-)} + 2\log{r_{\infty}^2}\,.
\end{align}

\subsection{$(1_{++},1_{+-})$} 
First let us put $A_2$ in Region $1_{+-}$.
This is what corresponds to the setup investigated in \cite{Hartman:2013qma}.
Let us take 
\begin{align}
\label{eq:PQansatz}
  \phi_2 - \phi_1 = \delta\phi\,,
\quad
  \ell_2 - \ell_1 = \delta\ell\,.
\quad
  t_2 - t_1 = \delta t\,,
\end{align}
Since the time coordinate $t$ flows to opposite directions 
between $1_{++}$ and $1_{+-}$ regions, 
we regard this $\delta t$ as the time flow of the total system.
From \tbref{tab:geodesics}, we obtain
\begin{align}
\label{eq:SAc+-}
S_A^{(c)} &=\log X_n^P + \log X_n^Q - 2\log(\pi^2T_+T_-) + 2\log{r_{\infty}^2}\,,
\nonumber\\
X_n^P&=
  \cosh(\pi T_-(\delta\phi-\delta t+2\pi n))　\nl
&\quad \times
  \cosh(\pi T_+(\delta\phi + \delta t+2\pi n))\,, \nonumber\\
X_n^Q&=
  \cosh(\pi T_-(\delta\phi-\delta t+\delta\ell+2\pi n)) \nl
&\quad \times
  \cosh(\pi T_+(\delta\phi+ \delta t+\delta\ell+2\pi n))\,.
\end{align}
Of course, when $T_- = T_+$, $\delta\phi=\delta\ell=0$ and $n=0$,
this reproduces the corresponding result in \cite{Hartman:2013qma}
(eq.(3.32)). 

Furthermore, one can show that 
\begin{align}
\label{eq:SAc_bound+-}
  S_A^{(c)} >
4\pi T_+\abs{\delta t} 
-4\log{2}
-2\log{(\pi^2T_+T_-)}
+2\log{r_{\infty}^2}
\,,
\end{align}
for arbitrary choice of $n$. 
Therefore in any cases,
$S_A^{(c)}$ becomes very large in proportion to $|\delta t|$, therefore 
$S_A^{(d)} < S_A^{(c)}$ and $S_A=S_A^{(d)}$ in late time.

In particular, in near-extremal case,
we find that the right-hand side 
also has a very large constant term $-2\log{T_+}$.
It corresponds to the divergence of the distance to the horizon in the extremal black hole, which can also be observed in the case of 5D non-rotating charged extremal black hole \cite{Andrade:2013rra}.
In terms of the boundary theory, it is closely related to the residual entropy, coming from IR degrees of freedom.
As a result, the disconnected phase is always favored and we experience no transition in the near-extremal setup.

\subsection{$(1_{++},1_{--})$} 
When we put $A_2$ in Region $1_{+-}$
in the same way as \eqref{eq:PQansatz},
we obtain from \tbref{tab:geodesics}
\begin{align}
\label{eq:SAc--}
S_A^{(c)} =&\log X_n^P + \log X_n^Q 
- 2\log(\pi^2T_+T_-) + 2\log{r_{\infty}^2}\,,
\nonumber\\
X_n^P=&
  -\sinh(\pi T_-(\delta\phi-\delta t+2\pi n)) \nl
&\quad \times
  \sinh(\pi T_+(\delta\phi+ \delta t+2\pi n))\,, \nonumber\\
X_n^Q=&
  -\sinh(\pi T_-(\delta\phi- \delta t+\delta\ell+2\pi n)) \nl
&\quad \times
  \sinh(\pi T_+(\delta\phi + \delta t+\delta\ell+2\pi n))\,.
\end{align}
As we noted at the end of the previous section,
these $X_n^P$ and $X_n^Q$ are not positive in general.
They tend to be positive in late time for fixed values of $n$,
but for any fixed time and other parameters, they become negative
by taking sufficiently large $n$.

To avoid this strange property of the periodicity, 
let us consider a decompactifying limit and ignore the windings (i.e., set $n=0$)%
\footnote{In \cite{Hartman:2013qma}, this limit is taken implicitly.
This can be explicitly given as the scale transformation of AdS$_3$, as
\begin{align}
  \phi = \Lambda^{-1}\ti{\phi}\,,
\qquad
  t = \Lambda^{-1}\ti{t}\,,
\qquad
  r = \Lambda\ti{r}\,,
\end{align}
where $\Lambda\to\infty$.
Then in terms of $\ti{\phi}$, the periodicity is $2\pi\Lambda\to\infty$.
Accordingly, the parameters of our black hole and subsystem $A$
are also written as
\begin{align}
  &r_{\pm} = \Lambda\ti{r}_{\pm}\,,
\quad
  T_{\pm} = \Lambda\ti{T}_{\pm}\,,
\quad
  r_{\infty} = \Lambda\ti{r}_{\infty}\,,
\nl
  &\phi_i = \Lambda^{-1}\ti{\phi}_i\,,
\quad
  t_i = \Lambda^{-1}\ti{t}_i\,,
\quad
  \ell_i = \Lambda^{-1}\ti{\ell}_i\,,
\end{align}
and we regard the tilded quantities as $\mathcal{O}(\Lambda^0)$.
This means a huge black hole in the bulk and tiny intervals on the boundary. 
We omit tildes hereafter.}. 
After fixing $n=0$, 
the lengths of the both geodesics are real
when $\delta t > \delta t_0 \equiv \max\{|\delta\phi|,|\delta\phi+\delta\ell|\}$.
We restrict the time in this regime
 and consider the time evolution of the entanglement entropy after $\delta t_0$.

At  $ \delta t \to \delta t_0$, $S_A^{(c)}$ is negatively divergent.
From there it increases monotonically along with $\delta{t}$,
and when $\delta t$ becomes large 
(i.e., $\delta t\gg T_+^{-1},|\delta\ell|,|\delta\phi|$),
\begin{align}
  S_A^{(c)}\simeq 2 {r}_+\delta {t} - {r}_-(2\delta {\phi}+\delta {\ell}) - 2\log(\pi^2 {T}_+ {T}_-) + 2\log {r}_\infty^2\,.
\end{align}
Therefore in this setup, we always experience a transition
from the connected phase to the disconnected one.

\subsection{$(1_{++},1_{--})$}
As we noted in section \ref{sec:geodesics}, the boundary of $1_{-+}$ is completely timelike to that of $1_{++}$,
and so 
it is not reasonable to consider the entanglement between $1_{++}$ and $1_{-+}$.

\section{Discussions}
\label{sec:discussion}

In this short letter, we discussed the entanglement in the pairs of
$(1_{++},1_{+-})$ or $(1_{++},1_{--})$ boundaries,
by computing the entanglement entropy of the union of two intervals $A_1$ and $A_2$.
In $(1_{++},1_{+-})$ case, we have two candidates for the minimal surfaces --- connected and disconnected ones ---, and we can also have a freedom of the winding $n$ around the periodicity \eqref{eq:periods}, for the connected surface. 
Phase transition between the two phases may or may not happen, depending on the parameters $T_{\pm},\delta\phi$ and $\delta\ell$.
In particular, in the near-extremal regime ($T_+\to 0$),
the disconnected phase is always favored and no transition takes place.
In $(1_{++},1_{--})$ case, the story is 
complicated because of the counterintuitive winding modes which contribute negatively to the 
spacelike distance.
After removing them by decompactification, we find that the phase transition always occurs.

We can 
also write down the entanglement entropies for $(1_{++},3_{\eta_1\eta_2})$ pairs.
However, the periodicity \eqref{eq:periods} makes problems again,
because it is clearly a closed timelike curve and so it is doubtful whether such sectors have physically consistent description as a field theory.
Furthermore, since the boundaries of region 3 are surrounded by the conical singularities (see \fgref{fig:penrose} (a)), we are not sure that we can rely on the standard prescription of the minimal area surface. 
The naive computation itself is an easy problem
by using \tbref{tab:analytic23},
and we leave it to the reader.

In this letter, we analyzed the relation between entanglement and multi-boundary connected spacetime in the three dimensional bulk. 
It would be interesting to generalize this to higher dimensional spacetime.   
For deeper understanding of how generic multi-boundary spacetime are emerging related to the boundary entanglement  
like \cite{Maldacena:2013xja}, 
we need to find a proper interpretation or counterparts of these results in the boundary CFT. 
Hopefully, we would return to these problems in near future.

\acknowledgements
This work was supported by RIKEN iTHES Project.  
NI is also supported 
in part by JSPS KAKENHI Grant Number 25800143.
NO thanks RIKEN Mathematical Physics Laboratory for hospitality 
while this work was being completed.
NO is also grateful to 
Kimyeong Lee, 
Futoshi Yagi, 
Zhaolong Wang, 
Sang-Jin Sin, 
Jae-Hyuk Oh, 
Shigenori Seki, 
Yunseok Seo 
and 
Yang Zhou 
for comments and discussions.

\appendix

\vspace{1cm}

\section{Spacetime Structure of Maximally Extended Rotating BTZ}
\label{app:BTZ}
In this appendix, we briefly review the spacetime structure of
the rotating BTZ black hole.
Large part of the contents here was examined in \cite{Hemming:2002kd},
and we use basically the same notation as theirs.

The AdS$_3$ spacetime is given as an
$\bR^{2,2}$-embedded hyperboloid, expressed by
\begin{align}
  \label{eq:hyperbolid}
  &x_0^2+x_1^2-x_2^2-x_3^2= R^2\,,\nonumber\\
  &ds^2 = -dx_0^2 - dx_1^2 + dx_2^2 + dx_3^2\,.
\end{align}
It is obvious that this space is invariant under 
$\SO(2,2)\simeq \SL(2,\bR)\times\SL(2,\bR)$,
and the AdS boundary is given by
\begin{align}
\label{eq:boundary}
  x_0^2 + x_1^2 \to \infty\,,
\qquad
  x_2^2 + x_3^2 \to \infty\,.
\end{align}
We take the AdS radius $R=1$ hereafter. 
By introducing $U$ and $V$ as
\begin{align}
\label{eq:UV}
  U= x_1^2-x_2^2\,,\qquad V= x_0^2-x_3^2\,,
\end{align}
the AdS hyperboloid \eqref{eq:hyperbolid} represents a straight line 
on the $(U,V)$-plane, 
\begin{align}
\label{eq:UVline}
  U+V=1\,.
\end{align}
At the same time, \eqref{eq:UV} can be regarded as hyperbolae
on $(x_1,x_2)$- and $(x_0,x_3)$-planes for each fixed pair $(U,V)$.
That is, each point $(U,V)$ on the line \eqref{eq:UVline} represents the direct product of a pair of these hyperbolae.
At $(U,V)=(1,0)$ and $(0,1)$, one of these two hyperbolae becomes a pair of
straight lines crossing at the origin. Note that from \eqref{eq:UVline}, we can decompose $(U,V)$-plane into three regions, 1: $U \ge 0, V \le0$, 2: $U  \ge 0, V \ge 0$, and 3: $U \le 0, V \ge 0$. This decomposition will be used later. 

In this context of \eqref{eq:UV}, the AdS boundary \eqref{eq:boundary} corresponds to
going to infinity on either (or both) of $(x_1,x_2)$- and $(x_0,x_3)$-planes
along with the hyperbolae.
Therefore obviously, every point $(U,V)$ on \eqref{eq:UVline} touches the AdS boundary.

The BTZ black hole \eqref{eq:BTZmetric} is obtained as an orbifold,
\begin{align}
  \label{eq:orbifold}
  \begin{pmatrix}
    x_1 \\
    x_2 \\
    x_3 \\
    x_0
  \end{pmatrix}
  &\simeq
  \begin{pmatrix}
    \cosh{\gamma_+} & \sinh{\gamma_+} & 0 & 0 \\
    \sinh{\gamma_+} & \cosh{\gamma_+} & 0 & 0 \\
    0 & 0 & \cosh{\gamma_-} & \sinh{\gamma_-} \\
    0 & 0 & \sinh{\gamma_-} & \cosh{\gamma_-}
  \end{pmatrix}
  \begin{pmatrix}
    x_1 \\
    x_2 \\
    x_3 \\
    x_0
  \end{pmatrix},\nonumber\\
  \gamma_{\pm} &= \pm 2\pi r_{\pm}\,,
\end{align}
of the global AdS$_3$ spacetime \eqref{eq:hyperbolid}.
This orbifolded spacetime can be covered by using 
$12$ patches, each of which has the metric of the form of \eqref{eq:BTZmetric}. Those are:
\begin{subequations}
\label{eq:embed1}
\begin{align}
\shortintertext{{\bf Region 1:} (outside the black hole, $r\ge r_{+}$.)}
  \label{eq:embed1a}
  x_1 &= \eta_1\l(\f{r^2-r_-^2}{r_+^2-r_-^2}\r)^{1/2}\cosh(\pi\l(T_+u^++T_-u^-\r)),\\
    \label{eq:embed1b}
  x_2 &= \eta_1\l(\f{r^2-r_-^2}{r_+^2-r_-^2}\r)^{1/2}\sinh(\pi\l(T_+u^++T_-u^-\r)),\\
  \label{eq:embed1c}
  x_3 &= \eta_2\l(\f{r^2-r_+^2}{r_+^2-r_-^2}\r)^{1/2}\cosh(\pi\l(T_+u^+-T_-u^-\r)),\\
    \label{eq:embed1d}
  x_0 &= \eta_2\l(\f{r^2-r_+^2}{r_+^2-r_-^2}\r)^{1/2}\sinh(\pi\l(T_+u^+-T_-u^-\r))\,,
\end{align}
\end{subequations}
Hereafter $u^{\pm}=\phi\pm t$, and
the pair $(\eta_1, \eta_2)$ takes $(+1,+1)$, $(+1,-1)$, $(-1,+1)$, $(-1, -1)$. 
This region 1 covers all the sign of $x_1$ and $x_3$ in the $(U, V)$-plane with $U>0$, $V \le 0$. 
\begin{subequations}
\label{eq:embed2}
\begin{align}
\shortintertext{{\bf Region 2:} (between the outer and inner horizons, 
$r_{-}\le r\le r_{+}$.)}
  \label{eq:embed2a}
  x_1 &= \eta_1\l(\f{r^2-r_-^2}{r_+^2-r_-^2}\r)^{1/2}\cosh(\pi\l(T_+u^++T_-u^-\r)),\\
  \label{eq:embed2b}
  x_2 &= \eta_1\l(\f{r^2-r_-^2}{r_+^2-r_-^2}\r)^{1/2}\sinh(\pi\l(T_+u^++T_-u^-\r)),\\
  \label{eq:embed2c}
  x_3 &= \eta_2\l(\f{r_+^2-r^2}{r_+^2-r_-^2}\r)^{1/2}\sinh(\pi\l(T_+u^+-T_-u^-\r)),\\
  \label{eq:embed2d}
  x_0 &= \eta_2\l(\f{r_+^2-r^2}{r_+^2-r_-^2}\r)^{1/2}\cosh(\pi\l(T_+u^+-T_-u^-\r)),
\end{align}
\end{subequations}
This region 2 covers $U\ge0$, $V \ge 0$ in the $(U, V)$-plane. 
Note that from region 1 to region 2, the range of $r$ changes from $r \ge r_+$ to $r \le r_+$, and the sign of $V = x_0^2 - x_3^2$ changes, 
while the sign of $U = x_1^2 - x_2^2$ unchanged.  This explains the $r-$dependent factor changes and 
the ``$\sinh$''-``$\cosh$'' flip between \eqref{eq:embed1c} and \eqref{eq:embed2c}, and between \eqref{eq:embed1d} and \eqref{eq:embed2d}. 
\begin{subequations}
\label{eq:embed3}
\begin{align}
\shortintertext{{\bf Region 3:} (inside the inner horizon, $r\ge r_{+}$.)}
  \label{eq:embed3a}
  x_1 &= \eta_1\l(\f{r^2-r_+^2}{r_+^2-r_-^2}\r)^{1/2}\sinh(\pi\l(T_+u^+-T_-u^-\r)),\\
  \label{eq:embed3b}
  x_2 &= \eta_1\l(\f{r^2-r_+^2}{r_+^2-r_-^2}\r)^{1/2}\cosh(\pi\l(T_+u^+-T_-u^-\r)),\\
  \label{eq:embed3c}
  x_3 &= \eta_2\l(\f{r^2-r_-^2}{r_+^2-r_-^2}\r)^{1/2}\sinh(\pi\l(T_+u^++T_-u^-\r)),\\
  \label{eq:embed3d}
  x_0 &= \eta_2\l(\f{r^2-r_-^2}{r_+^2-r_-^2}\r)^{1/2}\cosh(\pi\l(T_+u^++T_-u^-\r)).
\end{align}
\end{subequations}
This region 3 covers $U\le0$, $V > 0$ in the $(U, V)$-plane. 
 Note that region 1 and region 3 are related by $U (= x_1^2 - x_2^2)$ and $V (= x_0^2 -x_3^2)$ exchange, therefore,  
 region 1's $(x_1,x_2,x_3,x_0)$ and region 3's $(x_0,x_3,x_2,x_1)$ are exchanged%
\footnote{This explains relations between   \eqref{eq:embed1a} and \eqref{eq:embed3d},  \eqref{eq:embed1b} and \eqref{eq:embed3c},  \eqref{eq:embed1c} and \eqref{eq:embed3b}, and  \eqref{eq:embed1d} and \eqref{eq:embed3a}.}.

\begin{figure}
\centering
\begin{tabular}{ccc}
\includegraphics[width=20mm]{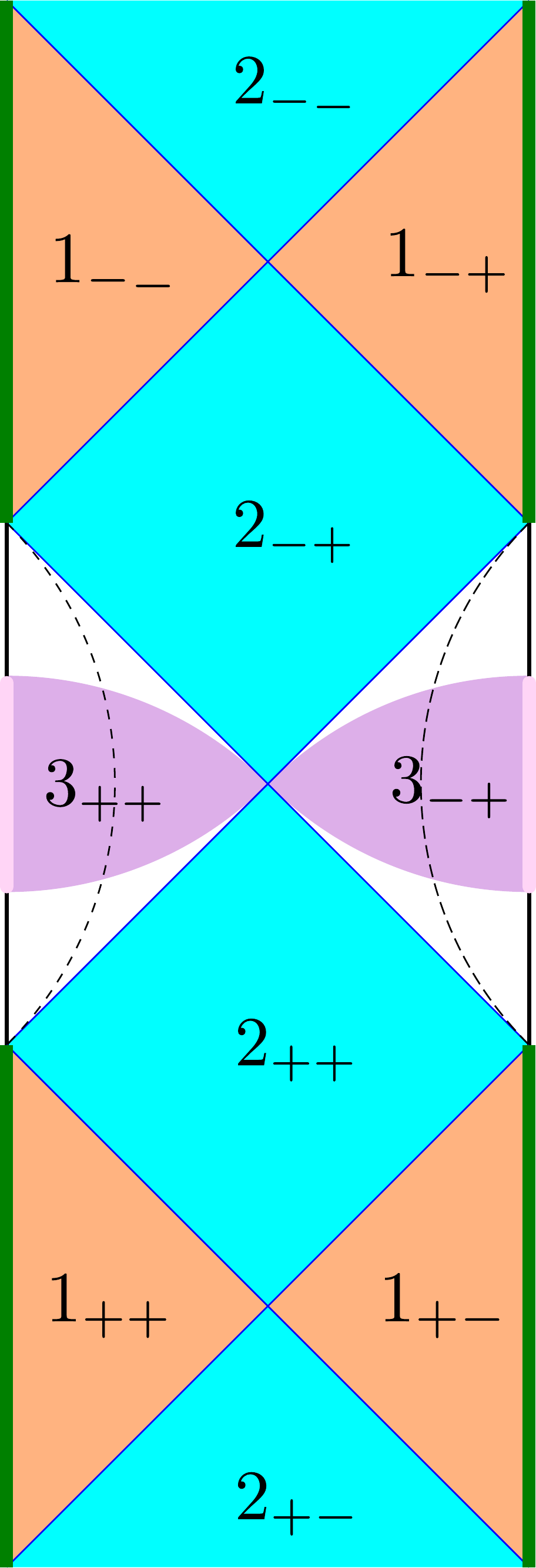}
&\hspace{5mm} &
\raisebox{2mm}{\includegraphics[width=55mm]{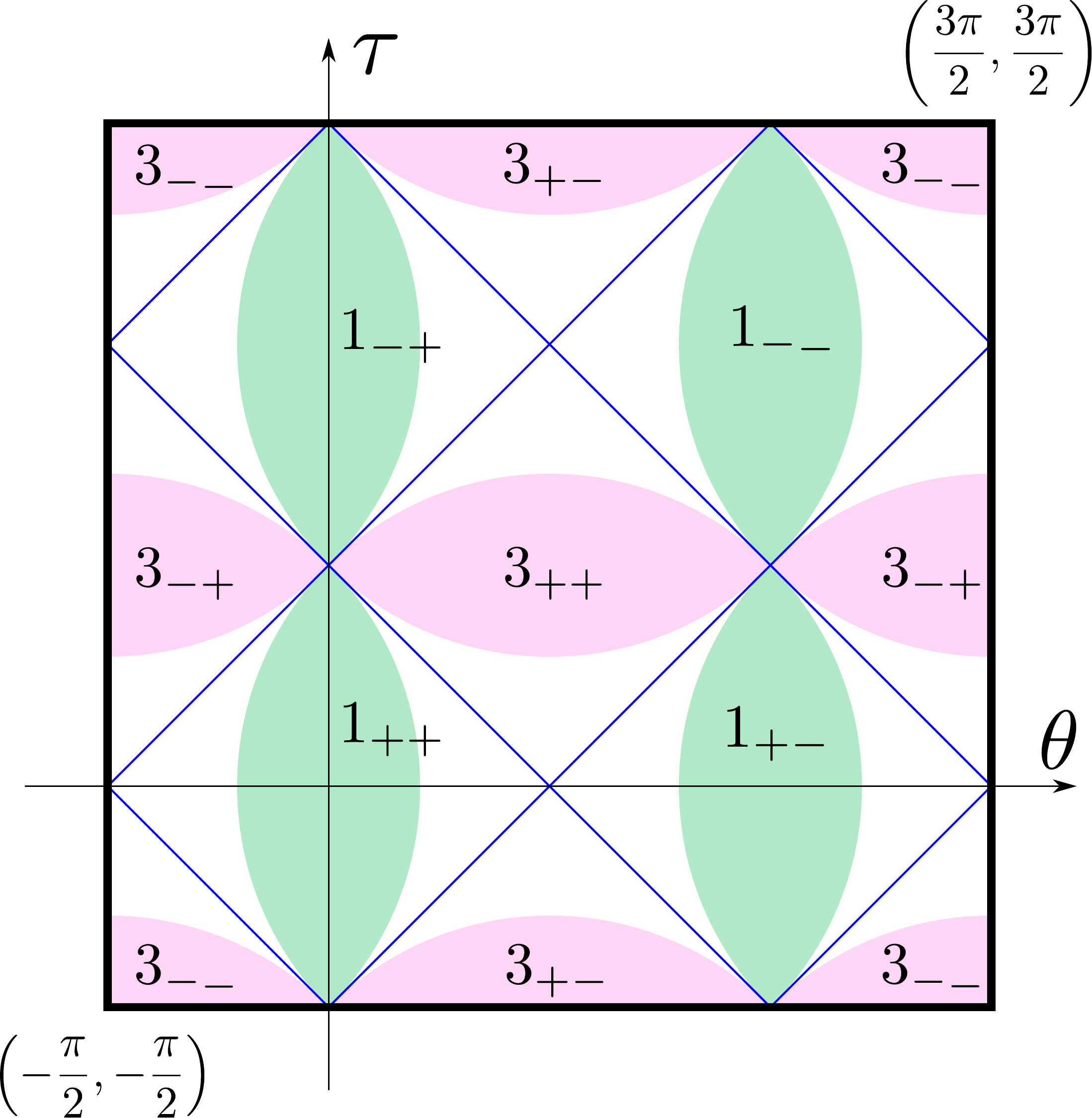}}
\\
(a) && (b)
\end{tabular}
\caption{
{\bf (a)}
The Penrose diagram of the rotating BTZ black hole with 
a two-dimensional $(t, r)$ plane set by $d\phi = N(r) dt$. 
Dashed lines represent BTZ (conical) singularity. 
\\
{\bf (b)}
The boundary in terms of global coordinate $(\theta, \tau)$, where $\theta$, $\tau$ have $2 \pi$ periodicity. 
BTZ identifications \eqref{eq:periods} restricts the fundamental domains as the colored areas. The diagonal blue lines represent region 2.\\
(These two figures are essentially copies of Fig.4 and Fig.5 in \cite{Hemming:2002kd}, respectively.)
}
\label{fig:penrose}
\end{figure}

Depending on these signs, we refer each of the $12$ regions as
$1_{++}$, $2_{+-}$, etc. 
It can be easily shown that each of the embeddings \eqref{eq:embed1}\eqref{eq:embed2}\eqref{eq:embed3} leads to the same induced metric \eqref{eq:BTZmetric},
while the orbifold \eqref{eq:orbifold} becomes
\begin{subequations}
\label{eq:periods}
\begin{alignat}{2}
  \phi&\simeq\phi+2\pi
  &\qquad& \text{(for region 1,2)}\,,\\
  t&\simeq t+2\pi 
  & & \text{(for region 3)}\,.
\end{alignat}
\end{subequations}
Due to the $ t \simeq t+2\pi$ identification in region 3, there is a  
conical singularity at  the 
radius $r = \sqrt{r_+^2 + r_-^2}$, where $g_{tt} = 0$ in region 3. 
The Penrose diagram for this spacetime
can be drawn as
\fgref{fig:penrose} (a)%
\footnote{Note that this diagram represents the null surface but the trajectory of the light 
is not necessary on this diagram, due to the constraint $d\phi = N(r) dt$.
}.

Since every point $(U,V)$ reaches to the AdS boundary, 
the AdS boundary is also divided into
the $12$ different regions $1_{\eta_1\eta_2}$, $2_{\eta_1\eta_2}$ and $3_{\eta_1\eta_2}$,
although the region 2 becomes just straight-lines on the boundary.
The arrangements of $1_{\eta_1\eta_2}$ and $3_{\eta_1\eta_2}$ on the AdS global coordinate boundary, 
can be seen, from 
$ \l.\tan\left(\frac{\theta \pm \tau}{2}\right)\r|_{\mu \to \infty}  = \l.\left( \tanh \left(\pi T_{\pm} u_{\pm}  \right) \right)^{\eta_1 \eta_2} \r|_{r \to \infty} $
for region 1, and 
$ \l.\tan\left(\frac{\theta \pm \tau}{2}\right)\r|_{\mu \to \infty} =  \l.\left( \pm \tanh \left(\pi  T_{\pm} u_{\pm} \right) \right)^{\mp \eta_1 \eta_2} \r|_{r \to \infty} $
for region 3.  The fact that $|\tanh \left(\pi u_{\pm}  T_{\pm}\right) | \le 1$ gives the 
 restriction for the allowed parameter range in the $(\theta, \tau)$ plane, and determines whether each boundary point belongs to region $1_{\eta_1\eta_2}$, or region $3_{\eta_1\eta_2}$. 
The configurations of each region on the AdS boundary is drawn in
\fgref{fig:penrose} (b).

\section{Analytic Continuations}
\label{app:analytic}
The different patches \eqref{eq:embed1}\eqref{eq:embed2}\eqref{eq:embed3}
can be connected to one another, by various analytic continuations of $(t,\phi,r)$ or $(u^{\pm},r)$ coordinates to complex-valued regions.
The list of the ones from $1_{++}$ to $1_{\eta_1\eta_2}$
is given in \tbref{tab:analytic1}. For completeness, we list the other 
formula of analytic continuations in \tbref{tab:analytic23}. 

\begin{table}
  \centering
  \begin{tabular}{c|c|c|c}
     & $u^+$ & $u^-$ & $r$\\
\hline
    $2_{++}$ & $u^+-\f{i}{4T_+}$ & $u^-+\f{i}{4T_-}$ & $r$\\
    $2_{--}$ & $u^++\f{3i}{4T_+}$ & $u^-+\f{i}{4T_-}$ & $r$\\
    $2_{+-}$ & $u^++\f{i}{4T_+}$ & $u^--\f{i}{4T_-}$ & $r$\\
    $2_{-+}$ & $u^+-\f{3i}{4T_+}$ & $u^--\f{i}{4T_-}$ & $r$\\
\newline
    $3_{++}$ & $u^+-\f{i}{2T_+}$ & $-u^-$ & $i\s{r^2-(r_+^2+r_-^2)}$\\
    $3_{--}$ & $u^++\f{i}{2T_+}$ & $-u^-$ & $i\s{r^2-(r_+^2+r_-^2)}$\\
    $3_{+-}$ & $u^+$ & $-u^--\f{i}{2T_-}$ & $i\s{r^2-(r_+^2+r_-^2)}$\\
    $3_{-+}$ & $u^+$ & $-u^-+\f{i}{2T_-}$ & $i\s{r^2-(r_+^2+r_-^2)}$
  \end{tabular}
  \caption{Analytic continuations from $1_{++}$ to $2_{\eta_1\eta_2}$ and $3_{\eta_1\eta_2}$,
up to the periodicity $(u^+,u^-)\simeq (u^++ i/T_+, u^-\pm i/T_-)$.
In region 2, we promise that $(r^2-r_+^2)^{1/2} = i(r_+^2-r^2)^{1/2}$.
}
  \label{tab:analytic23}
\end{table}

\bibliography{reference}

\end{document}